\documentclass[twocolumn,english,superscriptaddress, twocolumn,pre]{revtex4-1}
\usepackage[T1]{fontenc}
\usepackage[latin9]{inputenc}
\usepackage{geometry}
\usepackage{color}
\usepackage{verbatim}
\usepackage{mathtools}
\usepackage{amsmath}
\usepackage{amssymb}
\usepackage{graphicx}

\makeatletter

\providecommand{\tabularnewline}{\\}

\usepackage{amsmath}
\usepackage{subfigure}
\usepackage{graphicx, mathtools}
\usepackage[normalem]{ulem}
\usepackage[colorlinks=true,linkcolor=MidnightBlue,urlcolor=MidnightBlue,citecolor=MidnightBlue,anchorcolor=MidnightBlue]{hyperref}\usepackage[dvipsnames]{xcolor}

\usepackage{chngcntr}
\usepackage{listings}
\usepackage{xcolor, tikz-cd}
 
\definecolor{codegreen}{rgb}{0,0.6,0}
\definecolor{codegray}{rgb}{0.5,0.5,0.5}
\definecolor{codepurple}{rgb}{0.58,0,0.82}
\definecolor{backcolour}{rgb}{0.98,0.98,0.98}
 
\lstdefinestyle{mystyle}{
    backgroundcolor=\color{backcolour},   
    commentstyle=\color{codegreen},
    keywordstyle=\color{magenta},
    numberstyle=\tiny\color{codegray},
    stringstyle=\color{codepurple},
    basicstyle=\linespread{0.8}\footnotesize\ttfamily
    breakatwhitespace=false,         
    breaklines=true,                 
    captionpos=b,                    
    keepspaces=true,                 
    numbersep=5pt,                  
    showspaces=false,                
    showstringspaces=false,
    showtabs=false,                  
    tabsize=2
}

\makeatother

\usepackage{babel}
\begin{document}
\title{Ritz method for transition paths and quasipotentials of rare diffusive
events}
\author{Lukas Kikuchi}
\email{ltk26@cam.ac.uk }

\affiliation{DAMTP, Centre for Mathematical Sciences, University of Cambridge,
Wilberforce Road, Cambridge CB3 0WA, UK}
\author{Rajesh Singh}
\affiliation{DAMTP, Centre for Mathematical Sciences, University of Cambridge,
Wilberforce Road, Cambridge CB3 0WA, UK}
\author{M. E. Cates}
\affiliation{DAMTP, Centre for Mathematical Sciences, University of Cambridge,
Wilberforce Road, Cambridge CB3 0WA, UK}
\author{R. Adhikari}
\affiliation{DAMTP, Centre for Mathematical Sciences, University of Cambridge,
Wilberforce Road, Cambridge CB3 0WA, UK}
\affiliation{The Institute of Mathematical Sciences-HBNI, CIT Campus, Chennai 600113,
India}
\begin{abstract}
The probability of trajectories of weakly diffusive processes to remain
in the tubular neighbourhood of a smooth path is given by the Freidlin-Wentzell-Graham
theory of large deviations. The most probable path between two states
(the instanton) and the leading term in the logarithm of the process
transition density (the quasipotential) are obtained from the minimum
of the Freidlin-Wentzell action functional. Here we present a Ritz
method that searches for the minimum in a space of paths constructed
from a global basis of Chebyshev polynomials. The action is reduced,
thereby, to a multivariate function of the basis coefficients, whose
minimum can be found by nonlinear optimization. For minimisation regardless
of path duration, this procedure is most effective when applied to
a reparametrisation-invariant ``on-shell'' action, which is obtained
by exploiting a Noether symmetry and is a generalisation of the scalar
work {[}Olender and Elber, 1997{]} for gradient dynamics and the geometric
action {[}Heyman and Vanden-Eijnden, 2008{]} for non-gradient dynamics.
Our approach provides an alternative to chain-of-states methods for
minimum energy paths and saddlepoints of complex energy landscapes
and to Hamilton-Jacobi methods for the stationary quasipotential of
circulatory fields. We demonstrate spectral convergence for three
benchmark problems involving the Muller-Brown potential, the Maier-Stein
force field and the Egger weather model. 
\end{abstract}
\maketitle

\section{Introduction}

\textcolor{black}{The theory of Freidlin and Wentzell \citep{ventsel1970small}
gives asymptotic probability estimates of rare events in dynamical
systems perturbed by small noise}\textcolor{blue}{{} \citep{bolhuis2002transition,allen2005sampling,allen2009forward,ebener2019instanton}.}
Specifically, Freidlin-Wentzell theory yields estimates of the stationary
distributions and mean first-passage times. Both these quantities
are determined, in turn, by the asymptotic estimate of the probability
of a stochastic trajectory to not deviate from a smooth path by more
than a given amount in a given interval of time. The key result of
Freidlin and Wentzell is that the limiting form of this probability,
for small noise and small deviations, is given by a non-negative functional
of the smooth path. This functional is the Freidlin-Wentzell action
and its minimum, for fixed initial and terminal states, determines
both the stationary distributions and first-passage times. The smooth
path minimizing the action is often called the Freidlin-Wentzell instanton.
The theory is applicable to dynamical systems of both gradient and
non-gradient character and can so be used to study a wide variety
of equilibrium and non-equilibrium systems modelled by Itô diffusions
\citep{paninski2006most,huang2012molecular,bouchet2016generalisation,maier1996scaling,wolynes1995navigating,nolting2016balls,mangel1994barrier,gardner2000construction,demarco2001phase,nelson1987stochastic}.

Determining the minimum of the Freidlin-Wentzell action is a problem
in the calculus of variations. The Euler-Lagrange equations provide
the necessary conditions for extrema of variational problems and,
unsurprisingly, have been the basis of the large literature devoted
to the numerical computation of Freidlin-Wentzell instantons \citep{weinan2002string,paninski2006most,heymann2008geometric,grafke2017long}.
There exists, however, an alternative ``direct'' route for the solution
of variational problems in which the functional is reduced, through
finite-dimensional parametrizations of paths, to a multivariate function
and then extre\textcolor{black}{mised by appropriate multivariate
optimisation methods \citep{gelfand2012calculus,kantorovich1958approximate}.
To the best of our knowledge, the first use of the direct method for
the Freidlin-Wentzell action, discretised by finite-differences, appears
in the work of Weinan, Ren and Vanden-Eijnden \citep{weinan2004minimum}.} 

Here we combine the direct method with a Ritz discretisation \citep{ritz1909uber,gelfand2012calculus,kantorovich1958approximate}
to minimize the Freidlin-Wentzell action. We analyse paths in a spectral
basis of Chebyshev polynomials and use spectral quadrature to express
the action as a multivariate function of the basis coefficients. Nonlinear
optimisation is used to obtain coefficients that give the least action
from which the instanton is synthesised in the spectral basis.\textcolor{black}{{}
For minimisation over paths regardless of their duration, this procedure
is especially effective when applied to a reparametrisation-invariant
on-shell form of the action that follows from the time-translational
invariance of the Lagrangian. This generalises the scalar work functional
of Olender and Elber (for gradient dynamics) and the geometric action
of Heyman and Vanden-Eijnden} \citep{vanden2008geometric} (for non-gradient
dynamics). Our method is efficient enough to robustly sample the logarithm
of the asymptotic estimate of the stationary distribution, \emph{i.e.
}the quasipotential, avoiding the alternative, but numerically delicate,
route of solving the Hamilton-Jacobi equation \citep{cameron2012finding,yang2019computing,dahiya2018ordered}.
Our method is simple to use, converges rapidly, and is applicable
to both equilibrium and non-equilibrium problems. Its implementation
is freely available on GitHub as the open-source Python library PyRitz. 

The remainder of this paper is organized as follows. In the next section,
we recall key results of Freidlin-Wentzell theory from dual perspectives
of Itô stochastic differential equations and the corresponding Fokker-Planck
equations.\textcolor{black}{{} In Section (\ref{sec:Minima-of-the-action})
we present the derivation of the on-shell form of the Freidlin-Wentzell
action in a manner reminiscent of the Routh reduction procedure in
classical mechanics and explain its relation to the scalar work and
the geometric action. }In Section (\ref{sec:direct-method}) we describe
the direct method for the minimisation of functionals, the Chebyshev
spectral basis in which we construct smooth paths, the spectral quadrature
rule we use to evaluate the action, and the multivariate non-linear
optimisation methods we employ to find the minimum. In Section (\ref{sec:applications})
we apply the direct method to three well-known diffusion processes
and demonstrate convergence in each case. 

A particular achievement of our approach is its relatively facile
ability to calculate quasipotentials. This can be done with a sufficiently
high density of sample points to construct effectively continuous
maps of the quasipotential, which we do here for the same set of benchmark
problems. We conclude with a discussion on extending the method to
degenerate diffusion processes, systems with inertia and to the stochastic
dynamics of fields. 

\section{Large deviation theory\label{sec:fwtheory}}

We consider the autonomous dynamics of a $d$-dimensional coordinate
$X=(X^{1},\ldots,X^{d})$ in $\mathbb{R}^{d}$ perturbed by configuration-dependent
noise of intensity $\sqrt{\varepsilon}$ described by the Itô diffusion
equation

\begin{equation}
dX^{\mu}=a^{\mu}(X)dt+\sqrt{\varepsilon}\sigma_{\nu}^{\mu}(X)dW^{\nu}\label{general ito diffusion}
\end{equation}
governing the stochastic trajectory $X(t)$, where $a^{\mu}(X)$ is
the drift vector, $\sqrt{\varepsilon}\sigma_{\nu}^{\mu}(X)$ is the
volatility, $W^{\nu}(t)$ is a $d$-dimensional Wiener process and
repeated indices are summed over. The transition probability density
of the process, $P_{1|1}(x,t|x_{0})=P(X(t)=x|X(0)=x_{0})$, obeys
the Fokker-Planck equation $\partial_{t}P(x|x_{0})=\mathcal{L}P(x|x_{0})$
where the Fokker-Planck operator is

\emph{
\begin{equation}
\begin{aligned}\mathcal{L}(x) & =-\frac{\partial}{\partial x^{\mu}}a^{\mu}(x)+\frac{\varepsilon}{2}\frac{\partial^{2}}{\partial x^{\mu}\partial x^{\nu}}b^{\mu\nu}(x)\end{aligned}
\label{eq:fokker-planck}
\end{equation}
}and \textbf{$b^{\mu\nu}(x)=\sigma_{\lambda}^{\mu}(x)\sigma_{\sigma}^{\nu}(x)\delta^{\lambda\sigma}$}
is the diffusion tensor. We assume it to be non-degenerate, positive-definite
and invertible. The inverse, $b_{\mu\nu}(x)$, induces a Riemannian
structure in $\mathbb{R}^{d}$ with a norm $|x|_{b}=\sqrt{b_{\mu\nu}x^{\mu}x^{\nu}}$
that is distinct from the Euclidean norm $|x|=\sqrt{(x^{1})^{2}+\ldots(x^{d})^{2}}.$
We use the subscript $b$ to indicate this second ``diffusion''
norm. The stationary density satisfies the time-independent Fokker-Planck
equation\emph{ }$\mathcal{L}P_{1}(x)=0$ and, when it exists, is reached
asymptotically in time for arbitrary initial distributions, $\lim_{t\rightarrow\infty}P_{1|1}(x,t|x_{0})=P_{1}(x)$.

Associated with the Itô process is the Freidlin-Wentzell ``action''
functional \citep{ventsel1970small,graham1973statistical,graham1987macroscopic}
\begin{equation}
S[x(t)]=\frac{1}{2}\int_{0}^{T}|\dot{x}-a(x)|_{b}^{2}dt\label{eq:Freidlin-Wentzell action}
\end{equation}
which gives an asymptotic estimate for the logarithm of the probability
of trajectories $X(t)$ to remain in the tubular neighbourhood of
a smooth path $x(t)$ over the duration $0\le t\leq T$. We write
this as
\begin{equation}
P_{\text{tube}}[x(t)]\asymp\exp\left(-\frac{1}{\varepsilon}S[x(t)]\right)\label{eq:FW-LDP}
\end{equation}
which, in terms of limits, means

\[
S[x(t)]=\lim_{\delta\to0}\lim_{\varepsilon\to0}-\varepsilon\ln P\left[\sup_{0\leq t\leq T}|X(t)-x(t)|_{b}<\delta\right].
\]
The limits must be taken in the order above as they do not commute.
Eq. (\ref{eq:FW-LDP}) is a large deviation principle for trajectories
of Itô processes, due to Wentzell and Freidlin and Graham \citep{touchette2009large}.

For reasons described below, it is of interest to obtain the mode
of the tube probability over the set of continuous paths
\[
\gamma_{T}=\{x(t)\,|\,x(0)=x_{1},x(T)=x_{2},0\leq t\leq T)\}
\]
which have fixed termini $x_{1}$ and $x_{2}$ and are of duration
$T.$ This is equivalent to the variational problem of minimising
the Freidlin-Wentzell action. The minimum value of the action,

\begin{equation}
V_{T}(x_{2}|x_{1})=\min_{\gamma_{T}}S[x(t)],
\end{equation}
is called the quasipotential. The path attaining the minimum,

\begin{equation}
x_{T}^{*}(t)=\arg\,\min_{\gamma_{T}}\,S[x(t)],
\end{equation}
is called the instanton. We emphasise that this path describes the
smooth centerline of the tube of maximum probability and not a non-differentiable
trajectory of the diffusion process. It is the most probable dynamical
path connecting two points in configuration space. 

The instanton and the quasipotential are central objects in Freidlin-Wentzell-Graham
theory and relate to the eikonal approximation of the Fokker-Planck
equation \citep{ludwig1975persistence}. Assuming the JWKB form of
the transition density,
\[
P_{1|1}(x,t|x_{0})\sim\exp\left[\frac{1}{\varepsilon}\sum_{n=0}^{\infty}\varepsilon^{n}\phi_{n}(x,x_{0};t)\right],
\]
with prefactors suppressed, substituting in the Fokker-Planck equation
and matching terms gives a Hamilton-Jacobi equation for the lowest
order contribution,
\begin{equation}
\partial_{t}\phi_{0}+\frac{1}{2}b^{\mu\nu}\partial_{\mu}\partial_{\nu}\phi_{0}+a^{\mu}\partial_{\mu}\phi_{0}=0.
\end{equation}
This corresponds to the Hamiltonian system
\begin{align}
H(x,p) & =\frac{1}{2}b^{\mu\nu}p_{\mu}p_{\nu}+a^{\mu}p_{\mu}\\
\dot{x}^{\mu} & =+\frac{\partial H}{\partial p_{\mu}}=b^{\mu\nu}p_{\nu}+a^{\mu}\nonumber \\
\dot{p}_{\mu} & =-\frac{\partial H}{\partial x^{\mu}}=-\frac{\partial b^{\nu\lambda}}{\partial x^{\mu}}p_{\nu}p_{\lambda}-\frac{\partial a^{\nu}}{\partial x^{\mu}}p_{\nu}\nonumber 
\end{align}
whose solutions define an equivalent variational problem of extremising
an action with the Lagrangian

\begin{align}
L(x,\dot{x}) & =p_{\mu}\dot{x}^{\mu}-H(x,p)\label{eq:Legendre-form-of-Lagrangian}\\
 & =\frac{1}{2}(\dot{x}^{\mu}-a^{\mu})b_{\mu\nu}(\dot{x}^{\nu}-a^{\nu}).\nonumber \\
 & =\frac{1}{2}|\dot{x}-a(x)|_{b}^{2}\nonumber 
\end{align}
Thus, the rays of the Hamilton-Jacobi equation that determine the
lowest order contribution to the eikonal are local maxima of the tube
probability, or in other words, $\phi_{0}(x,x_{0};T)=V_{T}(x|x_{0})$.
The large-deviation principle of Freidlin and Wentzell and the theory
of the non-equilibrium potential of Graham \citep{graham1973statistical,graham1987macroscopic}
thus appear as elegant reformulations of the JWKB approximation \citep{ludwig1975persistence}. 

The correspondence with the JWKB approximation yields the asymptotic
form of the transition density,
\begin{equation}
P_{1|1}(x,T|x_{0})\asymp\exp\left[-\frac{1}{\varepsilon}V_{T}(x|x_{0})\right],
\end{equation}
and, in the $T\rightarrow\infty$ limit of the above, the asymptotic
form of the stationary distribution,
\begin{equation}
P_{1}(x)\asymp\lim_{T\to\infty}\exp\left[-\frac{1}{\varepsilon}V_{T}(x|x_{0})\right].
\end{equation}
If $x$ and $x_{0}$ belong to the same basin of attraction of an
attractor $\mathcal{A}$, then it can be shown that this limit is
independent of the initial coordinate,

\begin{equation}
\lim_{T\to\infty}V_{T}(x|x_{0})=V_{\infty}^{\mathcal{A}}(x),
\end{equation}
where $V_{\infty}^{\mathcal{A}}$ is equal, to within a constant,
to the stationary quasipotential $V_{\infty}(x)$ in the basin of
attraction of $\mathcal{A}$. For a system with multiple attractors
$\mathcal{A}_{i}$, the global quasipotential is

\begin{equation}
V_{\infty}(x)=\min_{i}\left(V_{\infty}^{\mathcal{A}_{i}}(x)+C^{\mathcal{A}_{i}}\right)\label{eq:aggregated quasi-potential}
\end{equation}
where $C^{\mathcal{A}_{i}}$ is an additive constant. The constants
are fixed by requiring

\begin{equation}
V_{\infty}^{\mathcal{A}_{i}}(x_{s}^{(i,j)})+C^{\mathcal{A}_{i}}=V_{\infty}^{\mathcal{A}_{j}}(x_{s}^{(i,j)})+C^{\mathcal{A}_{j}}
\end{equation}
for attractors $\mathcal{A}_{i}$ and $\mathcal{A}_{j}$ with adjacent
basins of attraction, where $x_{s}^{(i,j)}$ is the saddle with the
lowest value on the separatrix between the basins \citep{graham1987macroscopic}.
The stationary quasipotential determines the mean persistence time
of a trajectory in a basin of attraction which generalises the Arrhenius
law to systems out of equilibrium. 

The $T\rightarrow\infty$ limit involved in the definition of the
stationary quasipotential presents considerable numerical difficulties
in the minimisation of the Freidlin-Wentzell action. A more numerically
amenable route to determining the stationary quasipotential is by
the minimisation of the action over paths that start at an attractor
and end at a point in its basin, regardless of the duration. We show
in the next section that the solution of this second variational problem
does, indeed, yield the stationary quasipotential and derive an alternative
form of the Freidlin-Wentzell action that is adapted to computing
instantons regardless of their duration. %

\section{On-shell action\label{sec:Minima-of-the-action}}

We consider the variational problem of minimising the Freidlin-Wentzell
action over paths with fixed termini but of arbitrary duration,

\begin{equation}
\min_{T}\min_{\gamma_{T}}S[x(t)]=\min_{T}\min_{\gamma_{T}}\int_{0}^{T}L(x,\dot{x})dt,
\end{equation}
where both the initial and final points are in the basin of the attraction
$\mathcal{A}$ and the Freidlin-Wentzell Lagrangian following from
Eq. (\ref{eq:Legendre-form-of-Lagrangian}) is 

\begin{equation}
L(x,\dot{x})=\frac{1}{2}b_{\mu\nu}\dot{x}^{\mu}\dot{x}^{\nu}-b_{\mu\nu}a^{\mu}\dot{x}^{\nu}+\frac{1}{2}b_{\mu\nu}a^{\mu}a^{\nu}.
\end{equation}
This variational problem can be solved by introducing a parametrisation
$u$ for both the coordinate and time,

\[
x=x(u),\,\,x'=dx/du;\quad t=t(u),\,\,t'=dt/du,
\]
that allows the shape of the path, 

\begin{align*}
\sigma & =\{x(u)\,|\,x(u_{1})=x_{1},x(u_{2})=x_{2,}u_{1}\leq u\leq u_{2}\},
\end{align*}
to be varied independently of its duration, 
\[
T=\int_{u_{1}}^{u_{2}}t'du.
\]
Coordinates $x$ and time $t$ are dependent variables in the reparametrised
action,
\begin{equation}
S[x(u)]=\int_{u_{1}}^{u_{2}}L(x,\frac{x'}{t'})t'du,\label{eq:reparametrised-action}
\end{equation}
in which the time-dependence appears only through the derivative $t'$.
Therefore, $t$ is a cyclic (or ignorable) coordinate and Noether's
theorem implies that the corresponding conjugate momentum is conserved
\citep{whittaker1988}:
\begin{equation}
-\frac{\partial(Lt')}{\partial t'}=\frac{1}{2}b_{\mu\nu}\frac{x'^{\mu}x'^{\nu}}{(t')^{2}}-\frac{1}{2}b_{\mu\nu}a^{\mu}a^{\nu}=E.\label{eq:energy first integral}
\end{equation}
This defines submanifolds of the dynamics labelled by the ``energy''
$E$ which we shall call shells. The bound $2E+|a|_{b}^{2}\geq0$
for the energy follows immediately from the positive-definiteness
of the diffusion tensor. 

Solving the first integral for $t'$ gives
\begin{equation}
t'=\frac{dt}{du}=\frac{|x'|_{b}}{\sqrt{2E+|a|_{b}^{2}}},\label{eq:tprime}
\end{equation}
from which the duration of the path is obtained to be 

\begin{equation}
T_{E}=\int_{u_{1}}^{u_{2}}\frac{|x'|_{b}}{\sqrt{2E+|a|_{b}^{2}}}du.\label{eq:path T}
\end{equation}
This shows that paths $\gamma_{T_{E}}$ (of duration $T_{E}$) are
equivalent to shapes $\sigma_{E}$ (of energy $E$), where the latter
is the restriction of shapes in $\sigma$ to the shell of constant
energy. Then, minimisation over paths $\gamma_{T}$ regardless of
their duration is equivalent to minimisation over shapes $\sigma_{E}$
regardless of their energy, or
\[
\min_{T}\min_{\text{\ensuremath{\gamma_{T}}}}S[x(t)]=\min_{E}\min_{\text{\ensuremath{\sigma_{E}}}}S[x(u)].
\]
The action for shapes restricted to $\sigma_{E}$ is obtained by eliminating
$t'$ between Eq.(\ref{eq:reparametrised-action}) and Eq.(\ref{eq:tprime}).
This gives the ``on-shell'' form of the Freidlin-Wentzell action,
\[
S_{E}[x(u)]=\int_{u_{1}}^{u_{2}}\left[\frac{E+|a|_{b}^{2}}{\sqrt{2E+|a|_{b}^{2}}}|x'|_{b}-a^{\mu}x'_{\mu}\right]du,
\]
which is a functional of the shape $x(u)$, a function of the energy
$E$, and allows for independent variations of both. It is straightforward
to see that the integrand and, therefore, the action is minimised
when $E=0$. Therefore, most probable paths, regardless of their duration,
are obtained by minimising

\begin{equation}
S_{0}[x(u)]=\int_{u_{0}}^{u_{1}}\left(|a|_{b}|x'|_{b}-a^{\mu}x'_{\mu}\right)du
\end{equation}
over shapes restricted to the zero-energy shell. The duration on the
zero-energy shell,
\begin{equation}
T_{0}=\int_{u_{1}}^{u_{2}}\frac{|x'|_{b}}{|a|_{b}}du,
\end{equation}
shows that paths that leave, cross, or terminate at points of vanishing
drift, $a^{\mu}(x)=0$, are necessarily of infinite duration. The
corresponding shapes $\sigma_{0}^{\mathcal{\mathcal{A}}}$ can then
be taken to start at a fixed point and end at another point $x$ in
the basin of attraction. The quasipotential is determined by a minimisation
over such shapes $\sigma_{0}^{\mathcal{A}}$,

\begin{equation}
V_{\infty}^{\mathcal{A}}(x)=\min_{\sigma_{0}^{\mathcal{A}}}S_{0}[x(u)],
\end{equation}
and the shape attaining the minimum,
\begin{equation}
x_{\infty}^{\ast}(u)=\arg\min_{\sigma_{0}^{\mathcal{A}}}S_{0}[x(u)],
\end{equation}
is the stationary instanton. The time on the instanton path can be
obtained by integrating $t'=|x'|_{b}/|a|_{b}$. The utility of the
on-shell form of the action is that it provides the shape of the path
independently of its duration. The latitude of obtaining the shape
from a parametrisation over a finite interval, even for paths of infinite
duration, is extremely useful in numerical work. 

The on-shell action is related to, but distinct from, the Jacobi action
in mechanics \citep{landau1959classical,gantmachner}, which, following
a Routh reduction \citep{whittaker1988}, would in this case be
\[
\hat{S}[x(u)]=\int_{u_{1}}^{u_{2}}\left[\sqrt{2E+|a|_{b}^{2}}\,|x'|_{b}-a^{\mu}x'_{\mu}\right]du.
\]
\textcolor{black}{Though both the on-shell and Jacobi action agree
on the zero-energy shell, only the former supports the interpretation
as least action for non-zero energies. Furthermore, variations of
the on-shell action have to respect the on-shell condition Eq. \ref{eq:tprime}
(in other words, the solutions of its Euler-Lagrange equations does
not coincide with its extrema). On the other hand, the Jacobi action
can be varied using the standard Euler-Lagrange approach.}

\textcolor{black}{For gradient dynamics, that is $a^{\mu}=b^{\mu\nu}\partial U/\partial x^{\nu}$,
the on-shell action generalises the ``scalar work'' functional of
Olender and Elber \citep{olender1997yet} to non-zero energies and
configuration-dependent diffusion tensors. For non-gradient dynamics,
where the drift cannot be so expressed, the on-shell action generalises
the geometric action of Heyman and Vanden-Eijnden \citep{heymann2008geometric}
to non-zero energies. The non-zero energy shell $|\dot{x}|_{b}^{2}=|a|_{b}^{2}+E$
admits the most general path consistent with time-translation invariance,
in contrast to the zero-energy shell where the magnitude of the velocity
is always equal to that of the drift, $|\dot{x}|_{b}^{2}=|a|_{b}^{2}$.
Such general paths determine the quasipotential and the asymptotic
form of the transition density for finite times and will be examined
in detail in future work. Accordingly, we set $E=0$ below. The derivation
of the on-shell action only requires time-translation invariance of
the Lagrangian and not, as in \citep{olender1997yet,heymann2008geometric},
their positive-definiteness. Thus, it can be applied to stochastic
actions whose Lagrangians are not necessarily positive-definite, as
for example the Onsager-Machlup action \citep{onsager1953fluctuations,stratonovich1989some}.
We now describe the Ritz method by which we minimise actions. }

\section{Ritz method \label{sec:direct-method}}

The direct method in the calculus of variations consists of constructing
a sequence of extremisation problems for a function of a finite number
of variables that, in the passage to the limit of an infinite number
of variables, yields the solution to the variational problem. The
two main families of direct methods are finite differences and Ritz
methods \citep{ritz1909uber,kantorovich1958approximate,elsgolts1973differential,gelfand2012calculus}.
In the latter, the solution of the variational problem is sought in
a sequence of functions
\[
\varphi_{1}(t),\,\varphi_{2}(t)\,,\ldots\varphi_{n}(t),\ldots
\]
each of which satisfies end point conditions. The path is expressed
as a linear combination of these functions
\begin{equation}
x_{n}(t)=\alpha_{1}\varphi_{1}(t)+\ldots+\alpha_{n}\varphi_{n}(t)
\end{equation}
which transforms the action from a functional of the path into a function
of the expansion coefficients,
\begin{align}
S(\alpha_{1},\ldots,\alpha_{n}) & =\int_{0}^{T}L(x_{n},\dot{x}_{n})dt\label{general action}\\
 & =\int_{0}^{T}L\left(\sum_{i=1}^{n}\alpha_{i}\varphi_{i},\sum_{i=1}^{n}\alpha_{i}\dot{\varphi_{i}}\right)dt.\nonumber 
\end{align}
Action minimisation now becomes a search for a set of coefficients,
$\alpha_{i}^{\ast}$ such that $S(\alpha_{1}^{\ast},\ldots,\alpha_{n}^{\ast})<S(\alpha_{1},\ldots,\alpha_{n})$.
The necessary condition for this is the vanishing of the gradient,
\begin{equation}
\frac{\partial S}{\partial\alpha_{i}}=0\quad(i=1,2,\ldots n),
\end{equation}
which is the Ritz system of non-linear equations. Coefficients satisfying
these conditions can be obtained by non-linear optimisation. The $n$-th
approximation to the minimum action path, $x_{T}^{\ast}(t)$, and
the minimum value of the action, $S[x_{T}^{\ast}(t)]$, are obtained
from these values of the coefficients. It is generally the case that
this sequence of approximations converges to the minimum of the variational
problem as $n\rightarrow\infty$ \citep{gelfand2012calculus,kantorovich1958approximate}. 

The method, then, has three parts: first, the choice of basis functions
$\varphi_{i}(t)$; second, the quadrature rule that integrates the
Lagrangian to obtain the action as a function of the expansion coefficients;
and third, the optimisation that yields the coefficients at the minimum.
Since each part is only loosely dependent on the others, Ritz methods
come in many varieties \citep{gander2012euler}. Our choices are centered
around Chebyshev polynomials as described below. Approximation by
Chebyshev polynomials and their optimality for the purpose are described
in \citep{trefethen2000spectral,boyd2001chebyshev,trefethen2013approximation}. 

\emph{Basis functions: }We consider a path $x(u)$ that is a Lipschitz
continuous function of the parameter $u$ in the interval $[-1,1].$
Then, it is has an absolutely and uniformly convergent Chebyshev expansion,
\[
x(u)=\sum_{k=0}^{\infty}a_{k}T_{k}(u),\,\,\,\,a_{k}=\frac{2}{\pi}\int_{-1}^{1}\frac{x(u)T_{k}(u)}{\sqrt{1-u^{2}}}du
\]
where $T_{k}(u)$ are Chebyshev polynomials of the first kind and
the integral must be halved for $k=0$. A suitable sequence of paths
can be constructed from the first $n$ terms of this infinite series.
However, it is computationally more convenient, for reasons that will
be clear below, to construct the sequence from $n$-th degree polynomials
that interpolate the path at the $n+1$ Chebyshev points

\begin{equation}
u_{j}=-\cos(j\pi/n),\quad(j=0,1,\ldots n).
\end{equation}
The $n$-th degree interpolant can be expressed in standard form as
a sum of Lagrange cardinal polynomials $\ell_{j}(u)$ or as a linear
combination of Chebyshev polynomials,

\begin{equation}
x_{n}(u)=\sum_{j=0}^{n}\alpha_{j}\ell_{j}(u)=\sum_{k=0}^{n}c_{k}T_{k}(u).\label{eq:path-interpolation-expansion}
\end{equation}
The coefficients $c_{k}$ are aliased versions of the coefficients
$a_{k}$. Since the cardinal polynomials have the property

\[
\ell_{j}(u_{k})=\begin{cases}
1, & j=k\\
0, & \text{otherwise,}
\end{cases}\quad(j,k=0,\ldots,n),
\]
 $x_{n}(u_{k})=\alpha_{k}$, that is, the expansion coefficients $\alpha_{k}$
are path coordinates at the Chebyshev points. Expressing the entire
path in terms its discrete coordinates has the advantage that end
point conditions can be imposed by setting
\begin{equation}
\alpha_{0}=x(u_{0})=x_{0},\quad\alpha_{n}=x(u_{n})=x_{1}.
\end{equation}
Admissible paths of degree $n$ are, then, parametrised by the $n-1$
independent coefficients $\alpha_{1},\ldots,\alpha_{n-1}$. In contrast,
imposing end point conditions in series form leads to a numerically
inconvenient linear dependence between the coefficients $c_{k}$.
The derivative of the path is a polynomial of degree $n-1$ that can
be expressed in terms of the interpolant as
\begin{equation}
x_{n}'(u)=\sum_{j=0}^{n}\alpha_{j}\ell'_{j}(u)=\sum_{j=0}^{n}\beta_{j}\ell_{j}(u)
\end{equation}
with the two sets of expansion coefficients related by the Chebyshev
spectral differentiation matrix
\begin{equation}
\beta_{j}=D_{jk}\alpha_{k},\quad D_{jk}=\ell'_{k}(u_{j}).
\end{equation}
We use the barycentric form of the Lagrange polynomials \citep{hamming2012numerical}
\begin{equation}
\ell_{j}(u)=\frac{w_{j}}{u-u_{j}}\bigg/\sum_{k=0}^{n}\frac{w_{k}}{u-u_{k}}.
\end{equation}
with weights \citep{salzer1972lagrangian}
\[
w_{j}=\begin{cases}
\frac{1}{2}, & j=0\\
(-1)^{j}, & j=1,\ldots n-1\\
\frac{1}{2}\cdot(-1)^{n}, & j=n.
\end{cases}
\]
This form is both numerically stable and, costing no more than $O(n)$
operations, efficient to evaluate. \citep{berrut2004barycentric}. 

Chebyshev interpolants converge exponentially for analytic functions
and algebraically for functions with a finite number of derivatives.
More precisely, for an analytic path, $||x-x_{n}||=O(\rho^{-n})$
for some $\rho>1$ as $n\rightarrow\infty$. For a path with $\nu$
derivatives and $\nu$-th derivative of bounded variation $K$, $||x-x_{n}||=O(Kn^{-\nu})$
as $n\rightarrow\infty$. These estimates are in the supremum norm
$||a||$, that is, the maximum of the absolute value of $a$ in the
interval $[-1,1].$ In contrast, finite-difference methods can only
achieve polynomial, but never exponential, rates of convergence, even
for analytic paths \citep{trefethen2000spectral,boyd2001chebyshev}.

\emph{Quadrature}: To reduce the action to a multivariate function
of the coefficients it is necessary to evaluate the integral
\begin{equation}
S(\alpha_{1},\ldots,\alpha_{n})=\int_{-1}^{1}L(x_{n}(u),x'_{n}(u))du
\end{equation}
using a quadrature rule. \textcolor{black}{For instance, quadrature
at the Chebyshev points $u_{j}$ gives
\begin{align*}
S(\alpha_{1},\ldots,\alpha_{n}) & =\sum_{j=0}^{n}\omega_{j}L(x_{n}(u_{j}),x'_{n}(u_{j}))\\
 & =\sum_{j=0}^{n}\omega_{j}L(\alpha_{j},\beta_{j})\\
 & =\sum_{j=0}^{n}\omega_{j}L(\alpha_{j},D_{jk}\alpha_{k})
\end{align*}
where $\omega_{j}$ are the quadrature weights. However, standard
quadrature rules at this set of $n$ Chebyshev points, which integrate
a polynomial of degree less than or equal to $n$ exactly, will generally
be inaccurate. The reason is that the Lagrangian has polynomial degree
different from, and usually greater than, the polynomial degree of
the path. For instance, when $b_{ij}$ is a constant, the term quadratic
in the velocities has twice the polynomial degree of the path. Therefore,
if the Lagrangian is to be integrated accurately, the order of the
quadrature must be different from, and in general greater than, the
polynomial degree of the path.}

Therefore, we define a second set of $n_{q}>n$ Chebyshev points
\begin{equation}
v_{j}=-\cos(j\pi/n_{q}),\quad(j=0,1,\ldots n_{q})
\end{equation}
and interpolate the path at these points. This is done efficiently
by matrix multiplication with a $(n_{q}+1)\times(n+1)$ matrix

\begin{align}
x_{n}(v_{j}) & =\sum_{k=0}^{n_{q}}B_{jk}\alpha_{k},\\
x'_{n}(v_{j}) & =\sum_{k=0}^{n_{q}}B_{jk}\beta_{k},
\end{align}
whose elements are derived from the barycentric interpolant
\begin{equation}
B_{jk}=\frac{w_{k}}{v_{j}-u_{k}}\bigg/\sum_{l=0}^{n_{q}}\frac{w_{l}}{v_{l}-u_{l}}.
\end{equation}
The Lagrangian is evaluated at these second set of points after which
Clenshaw-Curtis quadrature \citep{trefethen2000spectral,boyd2001chebyshev}
is used to evalute the action,
\begin{align}
S(\alpha_{1},\ldots,\alpha_{n}) & =\sum_{j=0}^{n_{q}}\omega_{j}L(x_{n}(v_{j}),x'_{n}(v_{j}))\label{eq:quadrature-formula}\\
 & =\sum_{j=0}^{n_{q}}\omega_{j}L(B_{jk}\alpha_{k},B_{jk}\beta_{k})\nonumber \\
 & =\sum_{j=0}^{n_{q}}\omega_{j}L(B_{jk}\alpha_{k},B_{jk}D_{kl}\alpha_{l})\nonumber \\
 & \equiv\sum_{j=0}^{n_{q}}\omega_{j}L(B_{jk}\alpha_{k},C_{jk}\alpha_{k}).\nonumber 
\end{align}
As with interpolation, Clenshaw-Curtis quadrature converges exponentially
for Lagrangians that are analytic in $u$ and algebraically for Lagrangians
with a finite number $u$derivatives. Precise estimates are given
in \citep{trefethen2013approximation}. For fixed values of $n$ and
$n_{q}$, the matrices $B_{ij}$ and $C_{ij}$ in the above expression
are constant and can be precomputed and stored. The multiplications
require $O(nn_{q})$ operations, \textcolor{black}{and so there is
a linear cost, for fixed $n$, to increase the order of the quadrature.}
For Lagrangians of polynomial order $n_{L}$, the number of quadrature
points must be $n_{q}>(n+1)n_{L}.$ For nonpolynomial Lagrangians,
$n_{q}$ has to be chosen to ensure that the $n_{q}$-th Chebyshev
coefficient is suitably small. Well-defined procedures exist for the
adaptive truncation of Chebyshev series \citep{aurentz2017chopping}
but here we use a simple rule of thumb and set $n_{q}=10n$ leaving
the implementation of more efficient truncations to future work. \textcolor{black}{We
note that in the direct finite-difference method, introduced in \citep{weinan2004minimum},
the path is interpolated at uniformly spaced points by a quadratic
polynomial and the Lagrangian is integrated using the trapezoidal
rule. This combination can exactly evaluate the action for Lagrangians
that are at most quadratic polynomials.} %

\emph{Optimisation}: To minimise the action over the expansion coefficients
$\alpha_{1},\ldots,\alpha_{n-1}$ we use both gradient-free and gradient-based
algorithms. For gradient-free algorithms we provide Eq.(\ref{eq:quadrature-formula})
directly. For algorithms that require the gradient, the chain rule
gives
\begin{align}
\nabla_{\alpha_{i}}S & =\frac{\partial}{\partial\alpha_{i}}\left[\sum_{j=1}^{n_{q}}\omega_{j}L(B_{jk}\alpha_{k},C_{jk}\alpha_{k})\right]\\
 & =\sum_{j=1}^{n_{q}}\left[\frac{\partial L}{\partial x_{n}(v_{j})}B_{ji}^{\star}+\frac{\partial L}{\partial x'_{n}(v_{j})}C_{ji}^{\star}\right]\nonumber 
\end{align}
where $B_{ij}^{\star}=\omega_{i}B_{ij}$ and $C_{ij}^{\star}=\omega_{i}C_{ij}$.
These matrices, too, can be precomputed and stored and only the partial
derivatives of the Lagrangian need to be computed for given values
of the coefficients. For the examples presented below, we use \emph{NEWUOA}
\citep{powell2006newuoa} for gradient-free optimisation and \emph{SLSQP}
algorithm \citep{kraft1988software} for gradient-based optimisation,
both of which are implemented in the \emph{NLOPT} numerical optimisation
package \citep{johnson2014nlopt}. For non-equilibrium systems, instantons
lose smoothness when passing through fixed points. For such paths,
convergence is still achieved but at less than spectral rates. Spectral
convergence can be recovered if paths are evaluated piecewise, taking
care to isolate the points of derivative discontinuities. This is
feasible because fixed points are the only locations where Freidlin-Wentzell
instantons can lose smoothness \citep{graham1987macroscopic}. %

\section{Numerical results\label{sec:applications}}

In this section, we apply the Ritz method to three diffusion processes
that are widely used to benchmark rare event algorithms. The first
is overdamped Brownian motion in a complex energy landscape, the second
is overdamped Brownian motion under the influence of a circulatory
force, and the third is a model of the weather. All three models have
configuration-independent diffusion tensors for which it is not necessary
to distinguish between covariant and contravariant indices. Python
codes for each of these examples are freely available on GitHub. 
\begin{figure*}[t]
\begin{centering}
\includegraphics[width=0.94\linewidth]{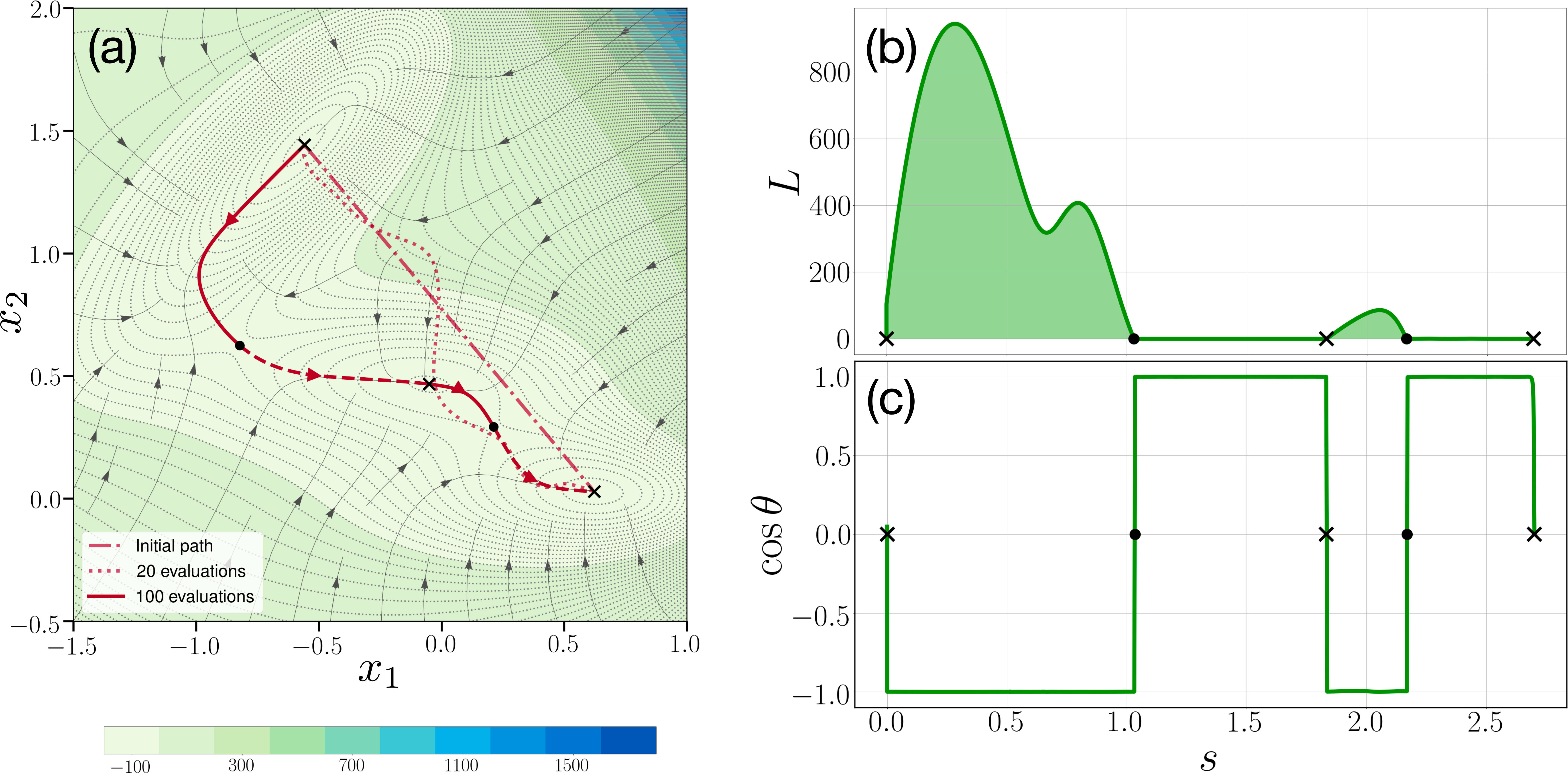}
\par\end{centering}
\caption{Ritz method for overdamped motion in the Muller-Brown potential, which
has three minima (crosses) and two saddle points (dots). The initial
path is the straight line connecting two minima and the instanton
is the solid line, with broken segments showing motion along the force.
The instanton automatically locates and passes through both saddles.
A typical path before convergence to the minimum is shown as a dotted
line. (b) The value of the Lagrangian as a function of Euclidean arc-length
of the instanton. The action vanishes to machine precision on segments
of the path where motion is along the force. (c) The cosine of the
angle $\theta$ between the tangent and the force is always $\pm1$,
i.e. the instanton is a minimum energy path. The instanton is represented
by a polynomial of degree $n=10$. }

\label{fig:muller-brown-instanton}
\end{figure*}

\subsection{Brownian dynamics in a complex potential}

Our first example considers the overdamped Brownian motion in a two-dimensional
potential with a constant friction. The usual equations of Brownian
dynamics can be recast into Itô form,

\begin{align*}
dX_{1} & =-\mu\partial_{1}Udt+\sqrt{2\mu\varepsilon}\,dW_{1}\\
dX_{2} & =-\mu\partial_{2}Udt+\sqrt{2\mu\varepsilon}\,dW_{2},
\end{align*}
where $\mu$ is the mobility and $\varepsilon=k_{B}T$ is the temperature.
The Freidlin-Wentzell action for a smooth path with two-dimensional
coordinate $x=(x_{1},x_{2})$ is
\[
S[x]=\frac{1}{2}\int_{0}^{T}\frac{1}{2\mu}|\dot{x}+\mu\nabla U|^{2}dt
\]
where $\nabla U=(\partial_{1}U,\partial_{2}U)$. The minimum of the
zero-energy action, 

\[
S_{0}[x]=\int_{-1}^{1}|\nabla U(x)||x'|du+\left[U(x)\right]_{-1}^{1},
\]
provides the most probable shape and the stationary quasipotential.
\textcolor{black}{The second term does not affect the minimisation
and can be discarded. The resulting reduced action}

\textcolor{black}{
\begin{equation}
\tilde{S}[x]=\int_{-1}^{1}|\nabla U||x'|du\label{eq:fermat}
\end{equation}
is of the same form as Fermat's principle for optical rays, where
$|\nabla U(x)|$ plays the role of the refractive index and $|x'|du=ds$
is the arc-length of the ray. In geometric optics, Fermat's principle
is equivalent to Huygen's principle and its ``wavelet equation''}

\textcolor{black}{
\begin{equation}
\partial_{i}U=|\nabla U|\frac{dx_{i}}{ds}.\label{eq:huygen}
\end{equation}
}This can be easily verified by differentiatiating it with respect
to arc-length, to obtain the eikonal equation

\textcolor{black}{
\[
\partial_{i}|\nabla U|=\frac{d}{ds}\left[|\nabla U|\frac{dx_{i}}{ds}\right],
\]
which is identical to the Euler-Lagrange equation of the zero-energy
action. The wavelet equation implies that the tangent $t=dx/ds$ to
the path is parallel to the gradient of the potential, or equivalently,
that rays are normal to contours of the potential. This is the well-known
condition for a minimum energy path and was first derived variationally
from the scalar work functional by Olender and Elber \citep{olender1997yet}.
It provides a stringent test of the fidelity of the paths }obtained
by minimisation. 
\begin{figure*}
\begin{centering}
\includegraphics[width=0.99\textwidth]{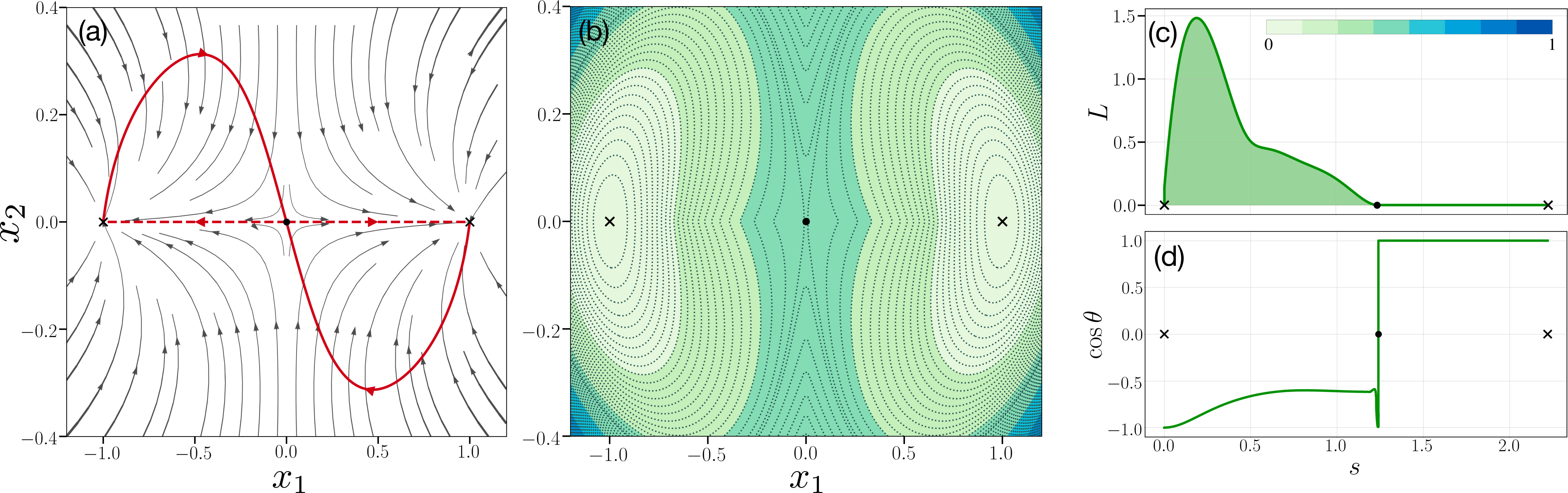}
\par\end{centering}
\caption{Ritz method for overdamped motion in a circulatory (i.e. non-gradient)
force field. The instanton is in red with solid (dashed) segments
showing motion against (along) the force field. The instanton is reflected
about the horizontal axis for motion starting on the right, showing
the inequivalence of fluctuational and relaxational paths for non-gradient
dynamics. (b) The quasipotential, computed using Eq. \ref{eq:aggregated quasi-potential},
with a caustic at the unstable fixed point. (c) The value of the Lagrangian
as a function of the Euclidean arc-length of the instanton. As in
the potential case, the action vanishes to machine precision on segments
where motion is along the force. (d) The cosine of the angle $\theta$
between the tangent and the force is, unlike in the potential case,
not always $\pm1$. The instanton is represented by a polynomial of
degree $n=8.$ }

\label{fig:maier-stein-quasipotentials}
\end{figure*}

Following \citep{olender1997yet}, we choose the Müller-Brown potential
of \citep{muller1979location} as an example of a complex energy landscape.
The potential and its stationary points are shown in Fig. \ref{fig:muller-brown-instanton}.
The three minima are marked by crosses and two saddle points by dots.
The instanton is computed by requiring the path to start at the minimum
on the top left and terminate at the minimum on the bottom right.
The initial straight line shape, an intermediate shape and the converged
instanton are shown in panel (a). The minimisation automatically locates
the two saddle points and makes the the instanton pass through them.
The action cost along the path is shown in panel (b), where the vanishing
of the action on segments of the path along the force is clearly seen.
The cosine of the angle between the tangent and force is shown in
panel (c) and the condition for a minimum energy path is clearly fulfilled.
We emphasise that the condition is not imposed separately but is satisfied
automatically at the minimum. The Ritz method provides an alternative
to chain-of-states methods for finding minimum energy paths. It does
not need the Hessian of the potential, which makes it suitable for
problems where such evaluations are expensive. Unlike \citep{heymann2008geometric},
our parametrisation has no unit-speed constraint and the minimisation,
accordingly, is unconstrained. The method applies without change to
dynamics with configuration-dependent friction. 
\begin{figure*}
\begin{centering}
\includegraphics[width=0.9\textwidth]{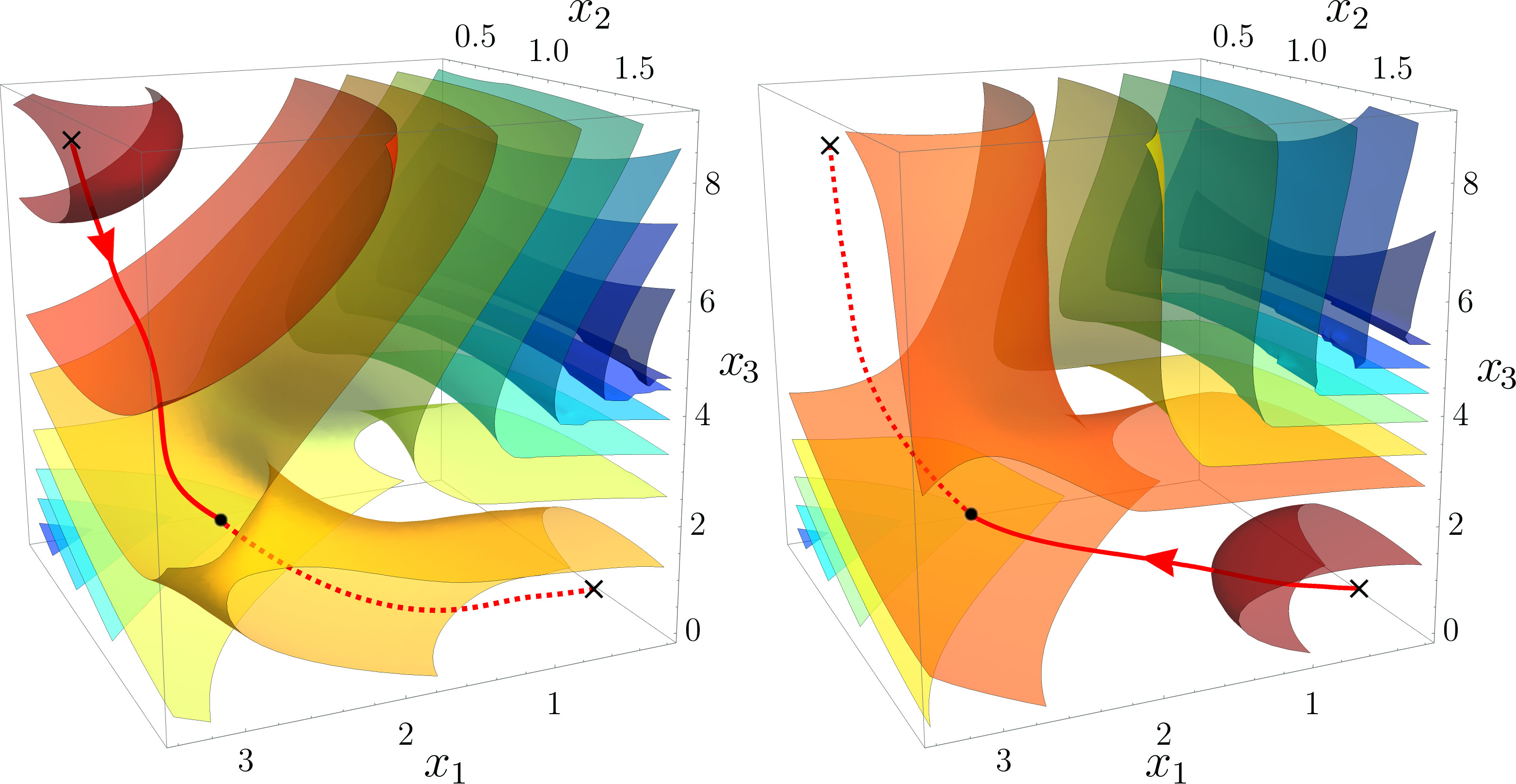}
\par\end{centering}
\caption{Instantons and quasi-potentials of the Egger model. The instantons
are shown in red with solid (dashed) lines representing motion against
(along) the vector field. The left and right panels are forward and
reverse instantons. Isosurfaces of the quasipotential with respect
to each attractor is shown in the respective panels. Isovalues increase
from light red to blue in the range $\left\{ 1,7,11,16,21,26,31,36\right\} $.
Parameter values are $k=2$, $\beta=1.25$, $\gamma=2$, $U_{0}=10.5$
and $H=12$. The instanton is represented by a polynomial of degree
$n=10.$}

\centering{}\label{fig:eggers}
\end{figure*}

\subsection{Brownian dynamics in a circulatory field}

Our second example consider, in contrast to the first, Brownian motion
in a force field that cannot be derived from a potential and, as such,
necessarily has a non-vanishing curl. Choosing the force field of
Maier and Stein \citep{maier1996scaling} gives

\[
\begin{aligned}dX_{1}= & (X_{1}-X_{1}^{3}-\beta X_{1}X_{2}^{2})dt+\sqrt{\epsilon}dW_{1}\\
dX_{2}= & -(1+X_{1}^{2})X_{2}dt+\sqrt{\epsilon}dW_{2}
\end{aligned}
\]
for the overdamped motion of the two-dimensional coordinate $X=(X_{1},X_{2})$,
where $\beta$ is a parameter. The force field $f(x_{1},x_{2})=(x_{1}-x_{1}^{3}-\beta x_{1}x_{2}^{2},-(1+x_{1}^{2})x_{2})$
is smooth, and $f_{1}$ is odd in $x_{1}$ and even in $x_{2}$, while
for $f_{2}$ the converse holds. There are two stable fixed points
at $x_{a}=(-1,0)$ and $x_{b}=(1,0)$, and a saddle point at $x_{s}=(0,0)$.
The force field admits a potential only for $\beta=1$, when it can
be written as $f=-\nabla U$, with $U(x_{1},x_{2})=-\frac{1}{2}x_{1}^{2}+\frac{1}{4}x_{1}^{4}+\frac{1}{2}(1+x_{1}^{2})x_{2}^{2}$.
The force field is shown in the first panel of Fig. \ref{fig:maier-stein-quasipotentials}
for $\beta=10$ together with the instanton moving from $x_{a}$ to
$x_{b}$. As before, solid (dashed) segments represent motion against
(along) the vector field. The instanton moving from $x_{b}$ to $x_{a}$
is obtained by reflection about the $x_{1}$-axis showing that that
fluctuational and relaxational paths are not identical in a non-gradient
field. 

The middle panels shows the stationary quasipotential $V_{\infty}^{\mathcal{A}_{i}}(x)$
with respect to the attractors at $(-1,0)$ and $(1,0)$ respectively.
The quasipotential is sampled on a $128\times128$ grid by computing
instantons between a point on the grid and the relevant attractor.
The contours of the quasipotential and its heatmap are obtained from
these discrete samples. To the best of our knowledge, all prior estimations
of the quasipotential for this problem (and more generally, for circulatory
forces) have required numerical solutions of the Hamilton-Jacobi equation.
Our method of direct sampling provides an alternative to this route
of computing the quasipotential. The right panel shows the Lagrangian
as a function of arc-length along the instanton. As in the previous
example, the Lagrangian vanishes along segments of the path where
motion is along the force. For motion against the force, the tangent
to the path is no longer parallel to the force, as shown by the variation
of the cosine of the angle $\theta$ between the tangent and the force.
We note that our method is agnostic to the existence, or not, of a
potential for the drift and treats both these cases on equal footing.
{} 

\subsection{Egger model of weather}

Our final example is a reduced model of the weather for a a three-dimensional
coordinate $X=(X_{1},X_{2},X_{3})$ that has a circulatory drift,
\begin{alignat}{1}
dX_{1}= & \left[kX_{2}\left(X_{3}-\frac{\beta}{k^{2}}\right)-\gamma X_{1}\right]dt+\sqrt{\epsilon}dW_{1}\nonumber \\
dX_{2}= & \left[kX_{1}\left(\frac{\beta}{k^{2}}-X_{3}\right)-\gamma X_{2}+\frac{HX_{3}}{k}\right]dt+\sqrt{\epsilon}dW_{2}\nonumber \\
dX_{3}= & \left[-\tfrac{1}{2}HkX_{2}-\gamma(X_{3}-U_{0})\right]dt+\sqrt{\epsilon}dW_{3}\label{eq:egger}
\end{alignat}
where $k$, $\beta$, $\gamma$, $U_{0}$ and $H$ are constants.
This model is due to Egger \citep{egger1981stochastically}. It is
not particularly illuminating to visualise the three-dimensional vector
field describing this dynamics but we note that it has two stable
fixed points, marked by crosses in Fig. \ref{fig:eggers}, and a saddle
fixed point marked by a dot. The instanton moving between these points
is shown as before in the left and right panels of the figure. Also
shown are isosurfaces of the quasipotential with respect to the stable
fixed points, with isovalues increasing from red to blue. To the best
of our knowledge, this is the first computation of the quasipotential
for this model. We provide this example primarily to demonstrate the
feasibility of sampling quasipotentials in dimensions greater than
two with our method. 
\begin{table*}
\begin{centering}
\begin{tabular}{|c|c|c|c|c|c|}
\hline 
Model & $S_{1}-S_{2}$ & $S_{3}-S_{4}$ & $S_{7}-S_{8}$ & $S_{15}-S_{16}$ & $S_{31}-S_{32}$\tabularnewline
\hline 
\hline 
M-B & $2$ & $3\times10^{-4}$ & $1\times10^{-7}$ & $5\times10^{-14}$ & $1\times10^{-13}$\tabularnewline
\hline 
M-S & $2\times10^{-3}$ & $3\times10^{-7}$ & $2\times10^{-12}$ & $1\times10^{-16}$ & $5\times10^{-16}$\tabularnewline
\hline 
Egger & $8\times10^{-3}$ & $1\times10^{-3}$ & $6\times10^{-7}$ & $4\times10^{-9}$ & $8\times10^{-13}$\tabularnewline
\hline 
\end{tabular}%
\begin{tabular}{|c|c|c|c|c|}
\hline 
$S_{2}-S_{50}$ & $S_{4}-S_{50}$ & $S_{8}-S_{50}$ & $S_{16}-S_{50}$ & $S_{32}-S_{50}$\tabularnewline
\hline 
\hline 
$4\times10^{-2}$ & $4\times10^{-5}$ & $8\times10^{-8}$ & $1\times10^{-12}$ & $3\times10^{-13}$\tabularnewline
\hline 
$2\times10^{-4}$ & $7\times10^{-11}$ & $2\times10^{-13}$ & $1\times10^{-16}$ & $2\times10^{-16}$\tabularnewline
\hline 
$5\times10^{-3}$ & $1\times10^{-4}$ & $1\times10^{-6}$ & $1\times10^{-8}$ & $2\times10^{-10}$\tabularnewline
\hline 
\end{tabular}
\par\end{centering}
\caption{Convergence of the action $S_{n}$ for a path of polynomial order
$n$. The abbreviations M-B and M-S refer to Brownian dynamics in
the Muller-Brown potential and the Maier-Stein force field respectively.
The first five columns show the difference $S_{n}-S_{n+1}$ while
the next five show the difference $S_{n}-S_{50}$. A tenth-order polynomial
typically gives at least six digits of accuracy. \textcolor{red}{\label{tab:Convergence}}}
\end{table*}

\section{Numerical convergence}

We briefly recall the convergence properties of the Ritz method, comprising
that of the basis functions, the quadrature, and the optimisation.
The Chebyshev interpolant is guaranteed to converge to the most probable
path, assuming that it is Lipschitz continuous, at a rate that increases
with the number of derivatives the path admits and is exponential
for a smooth path. Likewise, the Clenshaw-Curtis quadrature is guaranteed
to converge to the minimum of the action, assuming that the Lagrangian
is Lipschitz continuous. The optimal number of quadrature points for
accuracy to machine precision can be obtained by following the decay
of the Chebyshev coefficients of the Lagrangian and truncating at
that value beyond which the coefficients vanish to machine precision.
The optimisation has lesser theoretical guarantees than the interpolation
and quadrature, as is generally the case with search in high-dimensional
spaces. However, the residual of the Ritz system provides an empirical
measure for how closely the minimum has been located. In all three
examples (and in others not presented here) we have found both gradient-free
and gradient-based optimization to robustly locate the minima, and
gradient-based methods to yield faster convergence. We note that for
equilibrium problems, the gradient-free method does not require the
Hessian of the energy function, which can be of significant computational
advantage. In Table \ref{tab:Convergence} we show the spectral convergence
of the action with increasing polynomial order of the path for each
of our examples. 

\section{Discussion}

We have presented an efficient and accurate numerical method for computing
most probable transitions paths and quasipotentials of rare diffusive
events. The method directly minimises the Freidlin-Wentzell action
and thus provides an unified approach for transition paths in both
equilibrium and non-equilibrium systems. Our reparametrisation-invariant
form of the action, derived using a Noether symmetry, is well-suited
for numerical work and is a generalisation of the geometric action.
This frees us from the constraints of the commonly used arc-length
path parametrisation and offers the maximum flexibility in choosing
the space of polynomials in which action is minimised. Thus our method
is not limited to the Chebyshev polynomials in $[-1,1]$ used here
but easily admits trigonometric polynomials and, more generally, any
global basis. Numerical quadrature reduces the action to a multivariate
function of coefficients of the path polynomial whose minimum is obtained
by both gradient-free and gradient-based optimisation. This gives,
simultaneously, both the minimum value of the action and the most
probable path. This efficiency of the method allows us to repeatedly
compute minimum action paths between an attractor and a point in its
basin of attraction and, thereby, map out the quasipotential. The
quasipotential in a non-equilibrium steady state has the same significance
as the Gibbs distribution in equilibrium and our method provides a
robust way of obtaining it without the need to numerically solve the
Hamilton-Jacobi partial differential equation. 
\begin{figure}
\begin{center}
\begin{tikzcd}[column sep=small, row sep=small]      
& S[x] \arrow{ddl}[swap]{\delta} \arrow{ddr}{\mathcal{P}} &
\\     &  &
\\[2ex]     
\delta S[x]=0  & & S(\alpha) \hspace{0.5cm}
\end{tikzcd}
\end{center}  
\begin{center}
\vspace{-0.6cm}
\begin{tikzcd} 
\arrow{d}[swap]{\mathcal{P}} &     \arrow{d}{\nabla} 
\\[2ex]     
\text{Discrete E-L} & \text{Ritz system} 
\end{tikzcd} 
\end{center}  


\caption{Inequivalence of the direct and Euler-Lagrange routes to numerical
action minimization. Here, $\delta$ $\rightarrow$ functional variation,
$\mathcal{P}\rightarrow$ finite-dimensional projection, and $\nabla\rightarrow$
function minimisation. On the left branch, the action is first varied
to obtain the Euler-Lagrange equation and then projected onto a finite-dimensional
basis for numerical solution. On the right branch, the action is first
projected onto a finite-dimensional basis and then minimised to obtain
the Ritz system. The finite-dimensional projection of the Euler-Lagrange
equations is, in general, not identical to the Ritz system.}

\centering{}\label{fig:discrete-el-versus-ritz}
\end{figure}
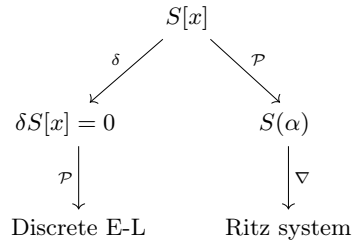

The direct method used here consists of a discretisation of the action
followed by a search for the minimum in the resulting finite-dimensional
space, expressed schematically in Fig. (\ref{fig:discrete-el-versus-ritz}).
In contrast, the majority of methods impose the vanishing variation
of the action and then search for the solution of the Euler-Lagrange
equation in a finite-dimensional space. The resulting discretised
Euler-Lagrange equations is, in general, not identical to the Ritz
system; in other words, these two methods of reducing an infinite-dimensional
problem to a finite-dimensional one are not equivalent. In contrast
to mechanics, where Newton's equations of motion are considered primary
and the action derived, here it is the tube probability and hence
the Freidlin-Wentzell action that is primary and the Euler-Lagrange
equation for the most probable path that is derived. It appears more
natural to us to discretise the primary, rather than the derived,
object directly. Our approach is algorithmically simple and the only
adjustable parameters are the polynomial order $n$ and the quadrature
order $n_{q}.$ This simplicity does not compromise accuracy or efficiency,
as confirmed by our examples. 

The rapid convergence of the method holds promise for its application
to problems involving the stochastic dynamics of fields, with both
scalar and Lie group-valued order parameters. We also expect the method
to apply to stochastic dynamics with degenerate diffusion tensors
and to stochastic systems with inertia. These will be addressed in
forthcoming work. 

\section{Acknowledgements}

We thank Julian Kappler for many helpful discussions. Work funded
in part by the European Research Council under the Horizon 2020 Programme,
ERC grant agreement number 740269. RS is funded by a Royal Society-SERB
Newton International Fellowship. MEC is funded by the Royal Society.

\end{document}